\documentclass[letterpaper, 10pt, conference]{ieeeconf}
\IEEEoverridecommandlockouts	

\usepackage{multirow}
\usepackage{lipsum}
\usepackage{amsmath}
\usepackage{amssymb}
\usepackage{subcaption}
\usepackage{graphicx}
\graphicspath{{./figures/}}
\usepackage{glossaries}
\usepackage[table,xcdraw]{xcolor}
\usepackage[ruled,vlined]{algorithm2e}
\usepackage[final]{listings}
\usepackage{longtable}
\usepackage[normalem]{ulem}
\usepackage{tikz,xcolor,hyperref}
\definecolor{lime}{HTML}{A6CE39}
\DeclareRobustCommand{\orcidicon}{
	\begin{tikzpicture}
	\draw[lime, fill=lime] (0,0) 
	circle [radius=0.16] 
	node[white] {{\fontfamily{qag}\selectfont \tiny ID}};
	\draw[white, fill=white] (-0.0625,0.095) 
	circle [radius=0.007];
	\end{tikzpicture}
	\hspace{-2mm}
}
\foreach \x in {A, ..., Z}{\expandafter\xdef\csname orcid\x\endcsname{\noexpand\href{https://orcid.org/\csname orcidauthor\x\endcsname}
			{\noexpand\orcidicon}}
}

\definecolor{mygray}{rgb}{0.5,0.5,0.4}
\definecolor{mygreen}{rgb}{0.2,0.5,0.2}
\definecolor{myblue}{rgb}{0.2,0.2,0.9}
\definecolor{mykwdclr}{rgb}{0.2,0.6,0.6}

\lstdefinestyle{mybase} {
	language=[Sharp]C,
	breaklines=true,
	showstringspaces=false,
	basicstyle=\small\tt,
	frame=single,
    numbers=left,
    numbersep=8pt,
	numberstyle=\tiny\color{mykwdclr},
	keywordstyle=\color{myblue},
	keywords=[2]{Mathf,Laser_Scanner,MonoBehaviour,Vector3,RaycastHit,Debug,Color,Physics}, 
	keywordstyle=[2]\color{mykwdclr}, 
	commentstyle=\color{mygreen}\ttfamily
}

\lstdefinestyle{mycsh} {
	style=mybase
}

\newacronym{ADAS}{ADAS}{Advanced Driver Assistance Systems}
\newacronym{ADS}{ADS}{Automated Driving Systems}
\newacronym{SAE}{SAE}{Society of Automotive Engineers}
\newacronym{TOR}{TOR}{Take Over Request}
\newacronym{DDT}{DDT}{Dynamic Driving Task}
\newacronym{ODD}{ODD}{Operational Design Domain}
\newacronym{NDRT}{NDRT}{Non-Driving Related Tasks}
\newacronym{LDS}{LDS}{Laser Distance Sensor}
\newacronym{LIDAR}{LIDAR}{Light Detection And Ranging}
\newacronym{RMSE}{RMSE}{root mean square error}

\title{\LARGE \bf Automated Driving Systems: Impact of Haptic Guidance on Driving Performance after a Take Over Request}

\author{Walter Morales-Alvarez$^{1}$\orcidD \emph{Student Member, IEEE}, Novel Certad$^{1}$\orcidF{} \emph{Student Member, IEEE}, \\ Hadj. Hamma Tadjine$^2$ \emph{Senior Member, IEEE} and Cristina Olaverri-Monreal$^{1}$\orcidE{} \emph{Senior Member, IEEE}%
\thanks{$^1$ Chair Sustainable Transport Logistics 4.0, Johannes Kepler University Linz, Altenberger Straße 69, 4040 Linz, Austria.
\texttt{\{walter.morales\_alvarez, novel.certad\_hernandez, cristina.olaverri-monreal\}@jku.at}}%
\thanks{IAV GmbH, Carnotstraße 1,  10587 Berlin, Germany.
\texttt{hadj.hamma.tadjine@iav.de}}%
}

\begin{document}

\maketitle
\thispagestyle{empty}
\pagestyle{empty}
\captionsetup[figure]{name={Fig.},labelsep=period}
\begin{abstract}
In conditional automation, a response from the driver is expected when a take over request is issued due to unexpected events, emergencies, or reaching the operational design domain boundaries. Cooperation between the automated driving system and the driver can help to guarantee a safe and pleasant transfer if the driver is guided through a haptic guidance system that applies a slight counter-steering force to the steering wheel. We examine in this work the impact of haptic guidance systems on driving performance after a take over request was triggered to avoid sudden obstacles on the road. We studied different driver conditions that involved Non-Driving Related Tasks (NRDT). Results showed that haptic guidance systems increased  road safety by reducing the lateral error, the distance and reaction time to a sudden obstacle and the number of collisions.

\end{abstract}

\section{Introduction}
\label{sec:introduction}
 

The integration of \gls{ADS} in the roads is expected to be gradual due to the technical complexity and social acceptance challenges that they involve~\cite{olaverri2020promoting}. In this context \gls{ADAS} with different levels of autonomy can support tactical and operational driving tasks~\cite{olaverri2017road}. 
At level 3 or conditional automation level, the \gls{ADS} of the vehicle can perform most of the \gls{DDT} under their \gls{ODD}, allowing users to engage in \gls{NDRT}. However, a response from the driver is still expected when a \gls{TOR} is issued due to unexpected events, emergencies, or reaching the \gls{ODD} boundaries~\cite{morales2020automated}. 
In this context, actions to take over the control of the vehicle require to take into account the drivers' cognitive load and/or attention level to the road as they will affect the time to respond to a \gls{TOR}. 
In line with this, \gls{ADS} and drivers can cooperate to guarantee a safe and pleasant transfer that diminish the effects of inattentive driving. For example, the \gls{ADS} can issue the \gls{TOR} through a warning buzzer while a haptic guidance system applies a slight counter-steering force to the steering wheel to guide the driver through the maneuver. 

To contribute to the state of the art, we examined in this work the impact of haptic guidance systems on driving performance after a \gls{TOR} was triggered. To this end, we defined a scenario in which sudden events forced drivers to perform maneuvers to avoid obstacles on the road. We also defined several automation scenarios and \gls{NDRT}s as independent variables to manipulate and measure the dependent variables that related to the driving performance. 

The  remainder  of  this paper  is  organized  as  follows: the next  section describes related  studies in  the  field;  section~\ref{sec:system_implementation_approach} details the experimental design; section~\ref{sec:evaluation}  presents  the  method  used to acquire and process the data collected; sections~\ref{sec:results} and \ref{sec:findings} present and discuss the obtained results. Finally section~\ref{sec:conclusion} concludes the present study outlining future research.

\section{Related Work}
\label{sec:relatedwork}

Several works have described the key factors that affect driving performance after having the driver regained the control of an automated vehicle. 
For example, results from driving simulation research showed that high traffic density decreased driving performance and increased the potential of maximum acceleration, leading to a lower time to avoid a collision~
\cite{du2020evaluating, 8594655}. \\
Safety can be increased through collaboration approaches in which shared control policies help the driver and \gls{ADS} to interact with each other~\cite{marcano2020review}. In this context, haptic guidance systems~\cite{steele2001shared} are often used
relying on driver's intention prediction approaches and modules that assess the decision of the percentage of control that the \gls{ADS} and driver have over the vehicle~\cite{erlien2013safe}. For example, in the work in~\cite{li2018shared}, authors used an inductive Multi-Label Classification with Unlabeled data (iMLCU) \cite{Wu2013MultiLabelCW} approach to classify drivers' intent, which was then compared to the desired maneuver
to establish the degree of control between the driver and the car. 

In an additional work, an architecture was proposed to calculate dynamic trajectories that took into account the driver's decisions. Actions to track the calculated trajectories using a control design were then performed relying on the Lyapunov method and Takagi-Sugeno Fuzzy model-based techniques~\cite{benloucif2019cooperative}.

Furthermore, studies like the one in~\cite{6775449} concluded that haptic guidance systems in straight and curve roads lead to a decrease in lateral error. 
Several degrees of system authority that ranged from no torque to a 100\% torque that was applied by the~\gls{ADS} were also studied in~\cite{6710125}, showing the results an increase in acceptability in scenarios with low degrees of shared control and low visibility.

Even though the works mentioned above designed and studied shared control policies to track specific paths, most of them only considered non-emergency scenarios. 
An exception constitutes the paper in~\cite{9517293}. However, a state of driver distraction through NDRT was not considered in this work. 
The authors studied how different shared control policies affected drivers' ability to avoid a sudden obstacle on the road. They found that driving performance increased as the system's authority over the vehicle increased, being the best results in situations in which the driver had no control over the vehicle. Another finding of the study was that high system authority yielded lower driver comfort. 

To the best of our knowledge, studies that analyze the influence of shared control on drivers that were performing NDRTs and are required to take over the control of the vehicle are scarce. Findings in the field very often refer to studies based on simulations, in which no drivers are involved. Therefore the resulting outcome is not applicable to real situations in which a control transfer to the driver is required.

Therefore, we contribute with this work to the body of research by assessing the impact of a haptic guidance system to avoid a sudden obstacle on the road after having a TOR been  issued and having been the drivers involved in secondary tasks. To this end, we implemented a haptic guidance system in the driver-centric simulation module of the 3DCoAutoSim simulation platform. \footnote{\small10}
\section{System Implementation Approach}
\label{sec:system_implementation_approach}


We implemented a haptic guidance system to investigate its effect on driving performance after the driver was requested to take over the control of the vehicle. To this end we defined several~\gls{NDRT}s and a scenario that forced drivers to avoid sudden obstacles on the road. We expected a certain outcome as a result of the effect of the system on the driving performance that we measured through the following metrics: maximum acceleration, lateral deviation, and distance to an obstacle on the road. We additionally measured the reaction time to start the obstacle avoidance maneuver from the time the obstacle appeared on the road. A detailed definition of these metrics is provided in Section~\ref{subsec:drivingparameters}.
We defined the potential outcome through the formulation of the following hypotheses:

\begin{itemize}
\item A)
H0. Haptic-guidance systems do not have any effect on drivers reaction time to perform an obstacle avoidance maneuver after a \gls{TOR} is issued.\\
H1.Haptic-guidance systems decrease drivers reaction time to perform an obstacle avoidance maneuver after a \gls{TOR} is issued.
\item B)
H0. Haptic-guidance systems do not have any effect on the lateral deviation after a \gls{TOR}.\\
H1. Haptic-guidance systems decrease lateral deviation after a \gls{TOR}.
\item C)
H0. Haptic-guidance systems do not have any effect on the minimum distance to an obstacle while performing an avoidance maneuver after a triggered \gls{TOR}. \\
H1. Haptic-guidance systems decrease the minimum distance to an obstacle while performing an avoidance maneuver after a triggered \gls{TOR}.
\item D)
H0. Haptic-guidance systems have no effect on the maximum acceleration applied by drivers after a \gls{TOR}. \\
H1. Haptic-guidance systems decrease the maximum acceleration applied by drivers after a \gls{TOR}.
\item E)
H0. Haptic-guidance systems do not have any effect on the number of drivers that collide with an obstacle after a \gls{TOR} is triggered.\\
H1. Haptic-guidance systems decrease the number of drivers that collide with an obstacle after a \gls{TOR} is triggered \gls{TOR}.
\end{itemize}



\subsection{Experimental setup and procedure}

\begin{figure*}
	\centering
	\includegraphics[width=0.8\textwidth]{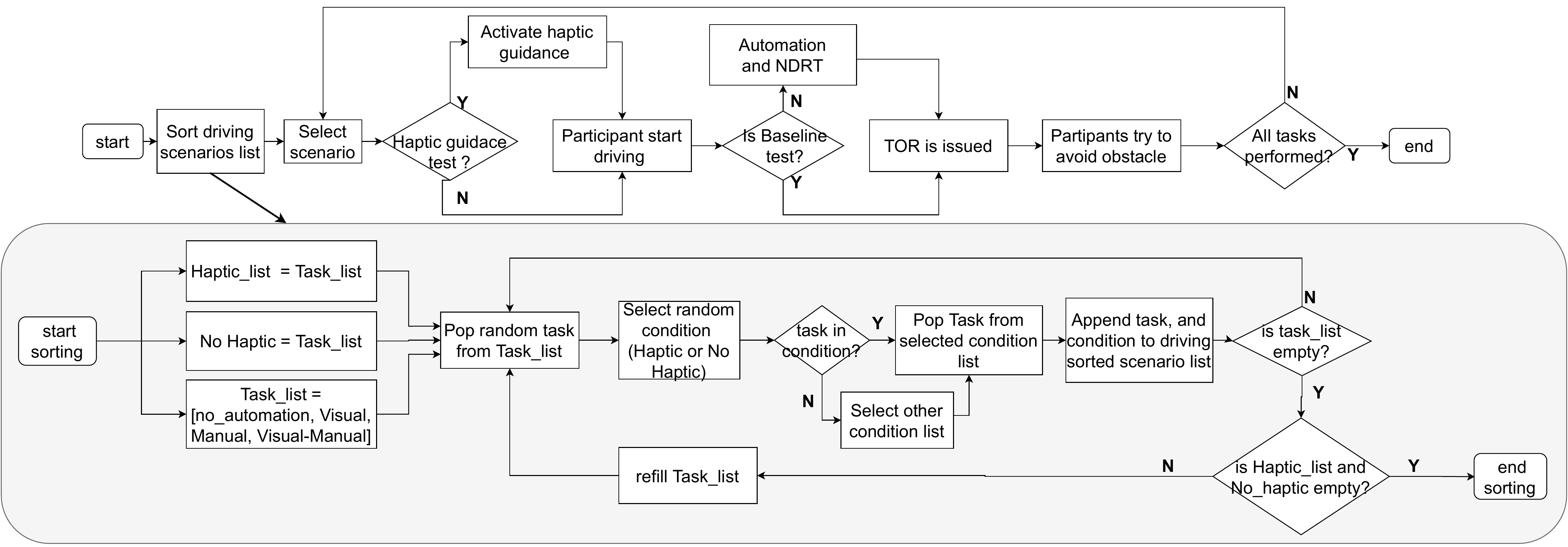}
	\caption{Experimental procedure flow diagram. The process to alternate the order of the different conditions was initiated through ``start sorting'', as illustrated within the graphic's red frame.}
	\label{fig:processflow}
\end{figure*}

Before starting the experiment, the participants were informed about the setup and instructed about how to engage and disengage the \gls{ADS} of the vehicle by pressing a button located in the steering wheel. To get familiarized with the simulator the participants were allowed to drive one lap through the road scenario. The total time to finalize the experiment was 30 minutes.
Previously to activating the automated driving mode, the drivers had to maintain their position on the right lane of the road until the system activated the automated mode. In automated mode the vehicle was programmed to follow the center of the right lane at 80 km/h and the drivers were asked to perform an \gls{NDRT} until a \gls{TOR} was issued. When an obstacle appeared on the road the driver needed to avoid it by changing lanes and then returning to the right lane. The order of the conditions was alternated to avoid bias as depicted in Figure~\ref{fig:processflow}.

\subsection{Driving scenario}
\begin{figure}
	\centering
	\begin{subfigure}{0.18\textwidth}
		\includegraphics[width=\textwidth]{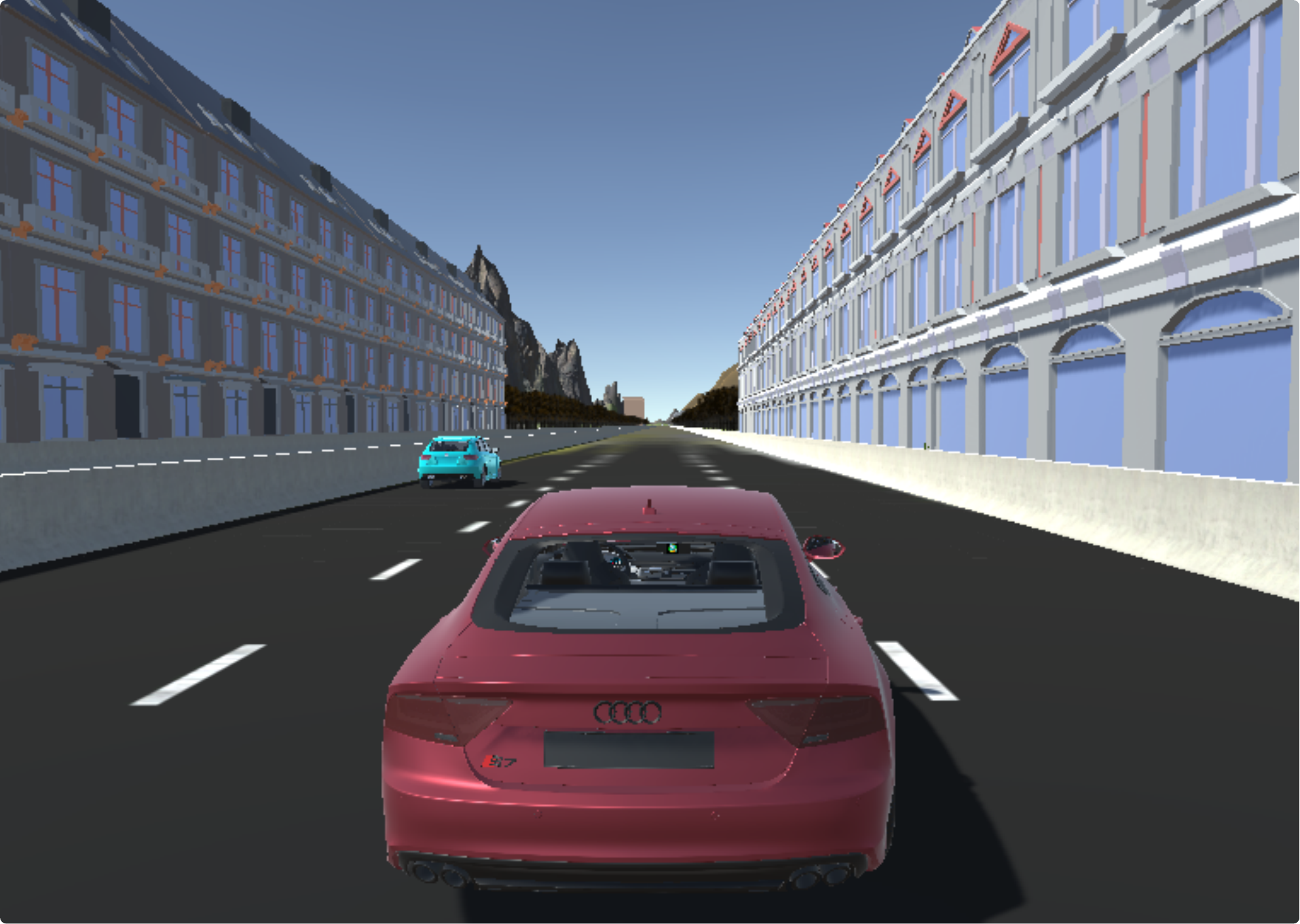}
		\caption{}
	    \label{fig:scene}
	\end{subfigure}
    \begin{subfigure}{0.17\textwidth}
		\includegraphics[width=\textwidth]{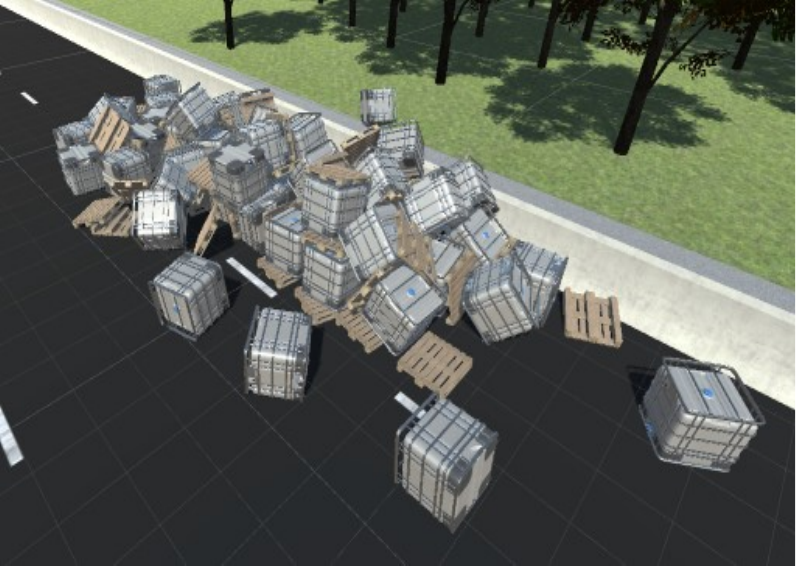}
		\caption{}
	    \label{fig:obstacle}
	\end{subfigure}
	\caption{Screenshots of the created scenario including (a) the designed road environment and (b) the obstacles that fell on the road.}
	\label{fig:unity}
\end{figure}

Figure \ref{fig:scene} shows the created scenario which consisted of a 15 km two-way highway (three lanes per way) with fences and trees on the side of the road. Mountains and buildings completed the scenario to provide the required realism. Along the road, a sudden event was triggered to replicate obstacles that can unexpectedly fall on the road (e.g. cargo from a truck or stones from the side of a mountain). They were located at a distance of 100 meters from the vehicle and consisted of a pile of boxes that fell on the road and spread through the driving lane, forcing drivers to avoid them and change the lane to continue driving (see Figure~\ref{fig:obstacle}). 

\subsection{Sample and simulation apparatus}

A sample of 23  participants with a valid driving license performed the simulation-based experiment (average age of 29.12 (SD = 15.16)). None of them had previous experience with \gls{ADS}. 

As previously mentioned, we used the driver-centric module of the 3DCoAutoSim simulation framework~\cite{8317937, 8569512}. This framework is a Unity3D based simulator, composed of a physical steering wheel, pedals, gear shift, and a comfortable car seat for a realistic driving experience. Three 4k monitors were installed in front of the steering platform to provide a wider field of view. The 3DCoAutoSim framework can simulate large 3D models, while allowing developers to add custom experimental scenarios and features to the framework~\cite{s21041523}.

\begin{figure}
	\centering
	\includegraphics[width=0.3\textwidth]{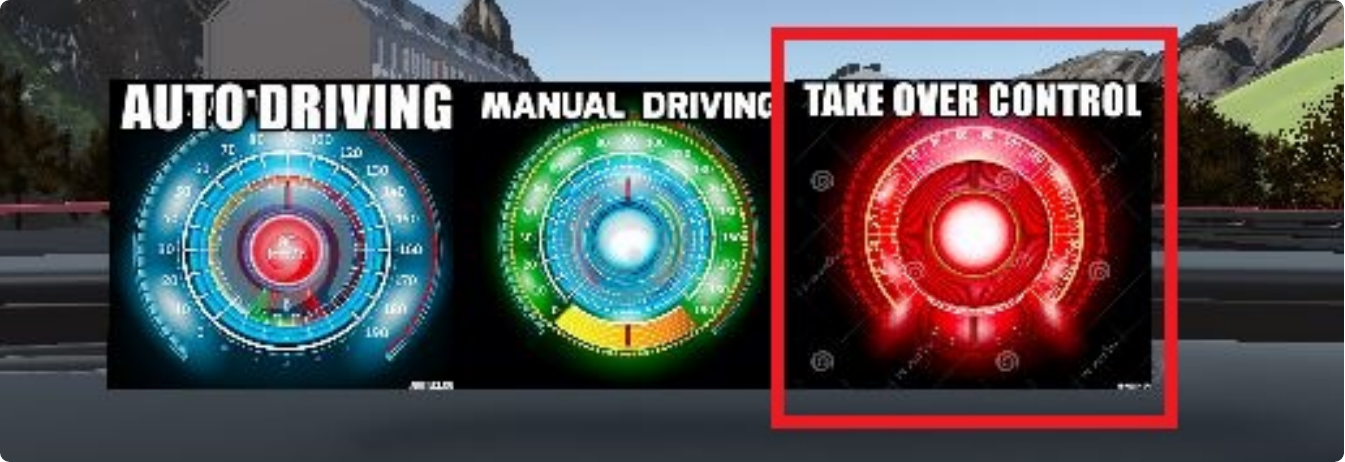}
	\caption{Visualization of the dynamic process to convey a TOR~\cite{olaverri2018automated}.}
	\label{fig:TOR}
\end{figure}


The \gls{NDRT}s were performed using a Samsung S10 mobile phone that the participants held while performing the tasks. These tasks were programmed in Unity to work under an Android phone. 

To collect the pertinent data we defined scenarios in which the participants were assisted by the haptic guidance system while  performing \gls{NDRT} in automated driving modus. To this end, we defined the system activation conditions and \gls{NDRT} as independent variables as detailed below.

\begin{enumerate}
  \item \textbf{Level 3:} The \gls{ADS} of the simulated vehicle was enabled. Participants were expected to take the control of the vehicle when a \gls{TOR} was issued to avoid the upcoming obstacle. The driver was requested to perform different \gls{NDRT}s. Driving without task in automated modus was evaluated as baseline condition.
    \item \textbf{Level 3 + haptic guidance:} The participants in the experiment regained control of the vehicle and avoided the obstacle with the assistance of the haptic guidance system when a \gls{TOR} was issued. Prior to this action they were performing \gls{NDRT}s. A baseline condition without any task was compared in this scenario, in which the \gls{ADS} cooperated with the participants by applying a small counter force to the steering wheel, helping them to avoid the obstacle on the road. 
\end{enumerate}
 
To evaluate the effect of cognitive workload and eyes off the road on the driving performance the following \gls{NDRT}s were performed relying on the field test presented in~\cite{morales2021real}:

\begin{itemize}
    \item \textbf{No Task:} No task was performed while the  vehicle was driven by the \gls{ADS}.
    \item \textbf{Visual:} The Stroop Color and Word Test (SCWT) \cite{jensen1966stroop}, was visualized on a provided mobile phone. The test consisted on speaking out loud the color in which a color name was written (e.g the work ``purple'' was written with a green font).
    \item \textbf{Manual:} Several M8x8 screws needed to be extracted from a bag filled with balls of 1 cm radius.
    \item \textbf{Visual-Manual:} A given text needed to be written backwards on the mobile phone. 
\end{itemize}
 
We manipulated these described conditions to study their effect on the dependent variables that related to driving performance and \gls{TOR} as described in section~\ref{sec:evaluation}.\\

\subsection{Haptic guidance system and TOR}
To perform the pertinent experiments we implemented a spring force feedback effect in the physical steering wheel of the simulator. We additionally implemented a custom software that served as interface between the simulation environment and the physical steering wheel. The torque applied to the steering was proportional to the angle between a given position and the current position of the steering. 

The Unity 3D object that triggered the sudden events issued at the same time a multimodal \gls{TOR}. This request consisted of a dynamic visual and acoustic signal of 480 Hz as described in~\cite{olaverri2018automated} that was conveyed on an in-vehicle display (see Figure~\ref{fig:TOR}).

\section{Data Collection and Evaluation}
\label{sec:evaluation}

We acquired driving-related data through the 3D CoAutoSim framework at a sampling rate of 10 Hz for the simulated GPS data, and 100 Hz for the simulated CAN data. 

\subsection{Driving parameters}
\label{subsec:drivingparameters}

We extracted and analyzed the following dependent variables:

\begin{itemize}
    \item \textbf{Reaction time (RT):} Defined as the elapsed time between the issue of the \gls{TOR} and the initiation of the obstacle avoidance maneuver~{\cite{doi:10.1177/0018720816634226}}. 
    \item \textbf{Lateral \gls{RMSE}:} Defined as root square mean of transverse error between the center of the lane and the position of the car \cite{8695742}. 
    \item \textbf{Distance to obstacle (DTO):} Defined as the minimum distance required to collide with an obstacle~\cite{ISO13586}. 
    \item \textbf{Maximum acceleration (MA):} Defined as the norm of the sum of the lateral and longitudinal acceleration when performing the obstacle avoidance maneuver~\cite{Gold2013TakeOH, kim2020understanding}.
    \item \textbf{Collisions:} Defined as the number of participants that collided with the upcoming obstacle for each scenario and secondary task.

\end{itemize}

\subsection{Data processing}

The acquired data was synchronized in the 3DCoAutoSim framework using a Unity's synchronization method between simulation processes. After this action, we calculated the maximum acceleration according to~\cite{Gold2013TakeOH}.

Given the longitudinal acceleration $a_{long}$ and lateral acceleration $a_{lat}$, the maximum acceleration is defined as:

\begin{equation}
    |a_{max}| = \sqrt{a_{lat}^2 + a_{long}^2}
\end{equation}

It is important to mention that the obtained accelerations were simulated and might therefore differ from real scenarios.

To calculate the lateral \gls{RMSE} after the issue of the \gls{TOR}, we estimated an optimal lane change path using heuristic methods. We then compared the position of the vehicle and the path to obtain the lateral error ($\Delta y_i$) during each time frame ($i$). The lateral \gls{RMSE} for $N$ sampled points was calculated by:

\begin{equation}
    RMSE = \sqrt{\frac{1}{N}\sum_{i=0}^{N}||\Delta y_i||^2}
\end{equation}

To determine the reaction time to start the obstacle avoidance maneuver, we triggered a timer along with the sudden event that caused the \gls{TOR}. The timer was stopped when the steering wheel was steered by at least 2 degrees~\cite{du2020evaluating, Gold2016TakingSituations}. According to the literature, this was the value that signalized that an obstacle avoidance maneuver was being performed in a scenario without sudden events. 

Finally, we obtained the minimum distance to obstacle ($DTO$) by the following means:

\begin{equation}
    DTO = \min{(||\Vec{r_t} - \Vec{r}_{obs}||)}
\end{equation}

Being $\Vec{r_t}$ the position of the vehicle at time step $t$ after the \gls{TOR} is triggered and $\Vec{r}_{obs}$ the corresponding obstacle location.

We additionally obtained the number of collisions by checking if the position of the vehicle $\Vec{r_t}$ over the path coincided with the bounding box $S_{xy} \in \mathbb{R}^2$ created by the obstacle, being $(r_{obs_x}, r_{obs_y})$ the $(x,y)$ position of the center of the obstacle, and $w_{obs}$ and $h_{obs}$ the respective width and height of the obstacle.

\begin{equation}
    Collision = \left\{
	\begin{array}{ll}
		true  & \mbox{if } \Vec{r_t} \in S_{xy}  \\
		false              & otherwise
	\end{array}
    \right.
\end{equation}

\begin{equation}
    S_{xy} = \left\{
	\begin{array}{ll}
		x & \forall x \in [r_{obs_x}-\frac{w_{obs}}{2},r_{obs_x}+\frac{w_{obs}}{2}]   \\
		y & \forall y \in [r_{obs_y}-\frac{h_{obs}}{2},r_{obs_y}+\frac{h_{obs}}{2}]
	\end{array}
    \right.
\end{equation}

\begin{figure*}[t]
    \centering
    \includegraphics[width=0.7\textwidth]{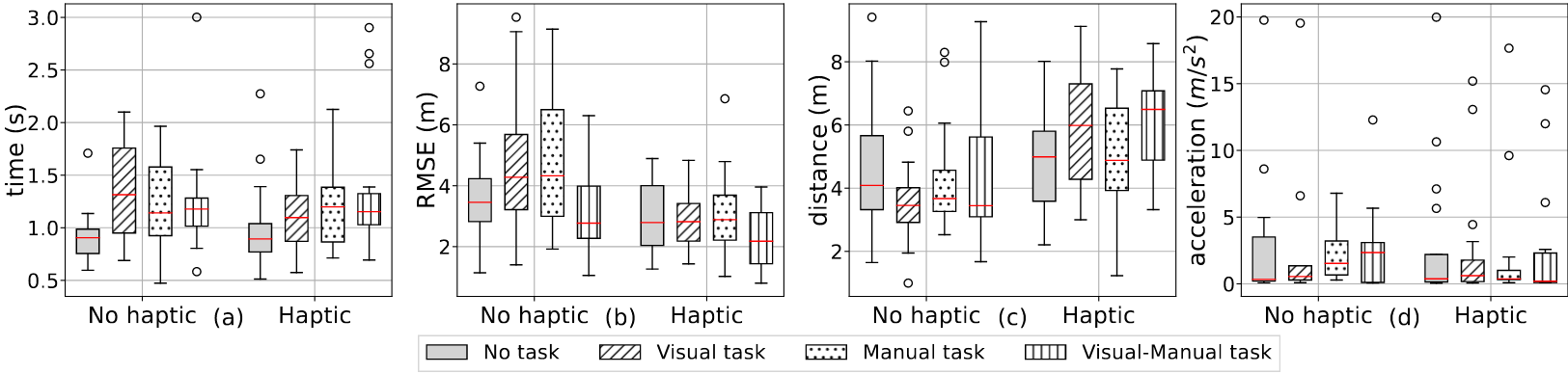}
    \caption{Visualization of the results: (a) Reaction time(RT); (b) minimum distance to collision object (DTO); (c) lateral root mean squared error (RMSE) and (d) maximum acceleration (MA)}
    \label{fig:bplots}
\end{figure*}

\begin{figure}[t]
    \centering
    \includegraphics[width=0.27\textwidth]{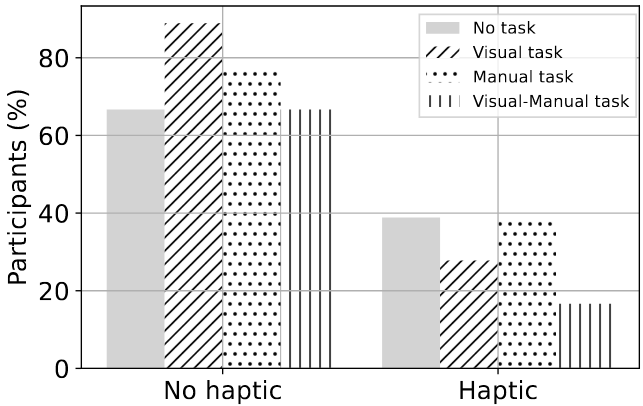}
    \caption{Visualization of the number of participants that collided with the obstacle.}
    \label{fig:hit}
\end{figure}

\begin{figure*}[t]
    \centering
    \includegraphics[width=0.6\textwidth]{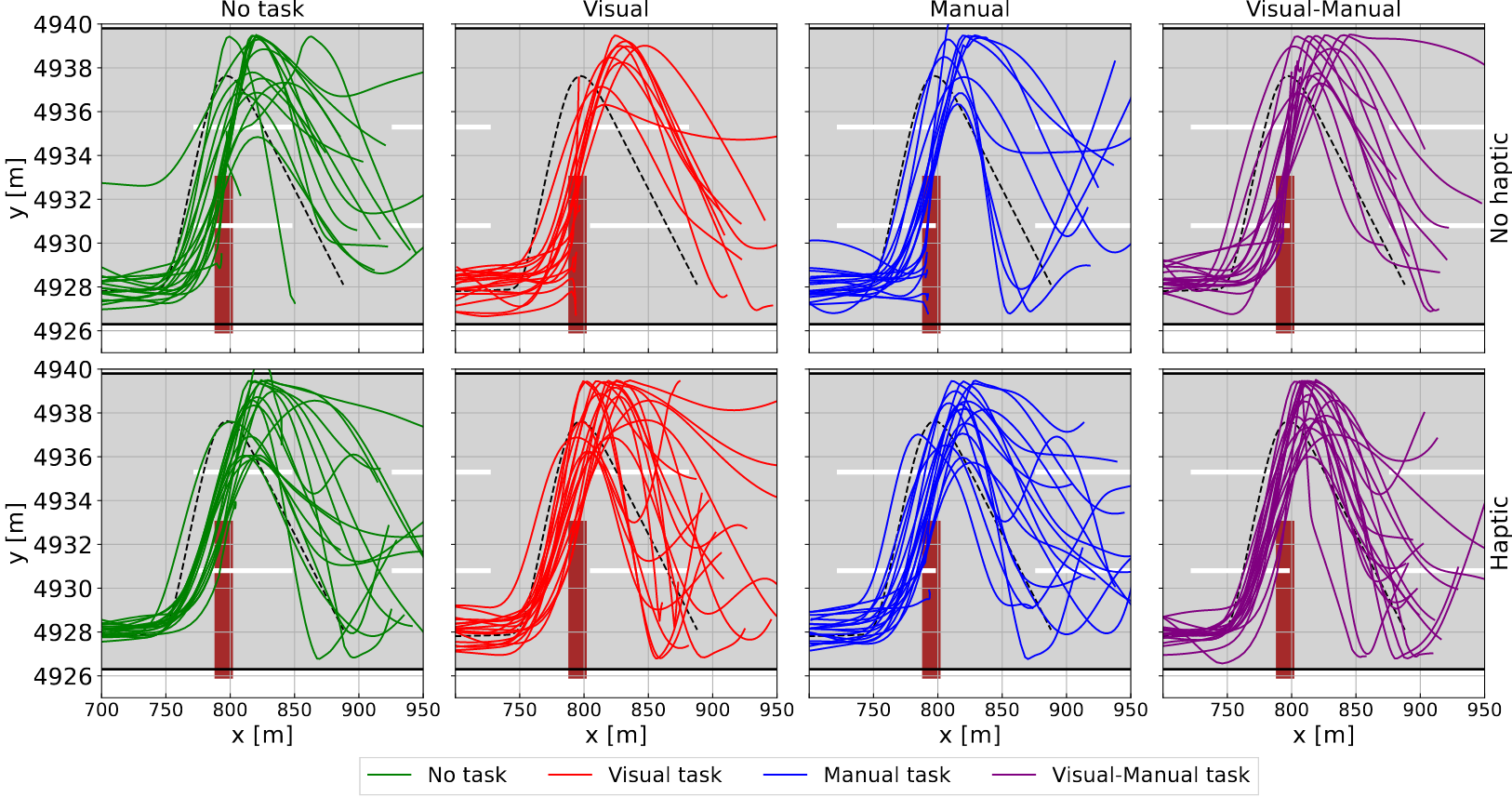}
    \caption{Visualization of the trajectories obtained for each participant after the \gls{TOR} was issued. From top to bottom the rows depict the path in scenarios without and with haptic guidance respectively. From left to right the columns depicts scenarios where no task, visual task, manual task and visual-manual task were performed. The brown rectangle represents in proportion the obstacle on the road.}
    \label{fig:paths}
\end{figure*}

\subsection{Statistical Analysis}

To analyze the acquired data and determine if there was a statistical significant relationship between the use of the haptic guidance system, secondary tasks variables and the dependent driving parameters, we performed a one Way ANOVA with a Bonferroni multiple post-hoc comparison.
We then compared the effect of using or not the haptic guidance system for each secondary task with a paired t-test. Additionally, as data was obtained from repeated samples, we performed a Cochran Q non-parametric test for dependent samples with a McNemar’s post hoc test to examine the statistical relationship between the use of the system and \gls{NDRT} on the number of drivers that collided with the obstacle for each test.


\section{Results}
\label{sec:results}

The results from the performed analyses when participants were exposed to a level 3 of automation with and without haptic guidance regarding reaction time (RT), lateral root mean square error (RMSE), minimum distance to obstacle (DTO), and maximum acceleration (MA) are depicted in Figure~\ref{fig:bplots} and Figure~\ref{fig:hit}. 

Figure~\ref{fig:paths} depicts the obtained trajectories that each participant traveled in the different scenarios. Table~\ref{table:mean_std} presents the mean and standard deviation values of each dependent variable at each scenario performed by the participants. 

Tables~\ref{table:mean_std} and~\ref{table:nohaptic_vs_haptic} present the statistical relationship between the dependent variables and the use / not use of the haptic guidance system after a take over request for each secondary task. Table~\ref{table:anova_cochran} shows if any particular~\gls{NDRT} benefits more from the activation of the haptic guidance system.

\begin{table*}
 \centering
 \scriptsize
 \caption{
    Mean and standard deviation regarding reaction time (RT), lateral root mean square error (RMSE), minimum distance to obstacle (DTO), maximum acceleration (MA). The number of collisions (Cols) is depicted at the bottom.}
 \label{table:mean_std}
 \begin{tabular}{|p{0.75cm}|p{3.71cm}|p{3.705cm}|p{3.705cm}|p{3.705cm}|}
  \hline
  Metric&No task&Visual task&Manual task& Visual-Manual task\\
 \end{tabular}\\
 \begin{tabular}{|p{0.75cm}|p{1.64cm}|p{1.63cm}|p{1.64cm}|p{1.64cm}|p{1.63cm}|p{1.64cm}|p{1.63cm}|p{1.64cm}|p{1.63cm}}
  \hline
  &No haptic&Haptic&No haptic& Haptic&No haptic& Haptic&No haptic& Haptic\\
 \end{tabular}\\
 \begin{tabular}{|p{.75cm}|p{0.6cm}|p{0.6cm}|p{0.6cm}|p{0.6cm}|p{0.6cm}|p{0.6cm}|p{0.6cm}|p{0.6cm}|p{0.6cm}|p{0.6cm}|p{0.6cm}|p{0.6cm}|p{0.6cm}|p{0.6cm}|p{0.6cm}|p{0.6cm}|}
  \hline 
  &$M$&SD&$M$&SD&$M$&SD&$M$&SD&$M$&SD&$M$&SD&$M$&SD&$M$&SD\\
  \hline
  RT& 0.907 & 0.248 & 1.001 &0.402&1.358&0.469&1.086&0.290&1.225&0.421 & 1.221 & 0.411 &1.225&0.482&1.368&0.623\\
  \hline
  RMSE& 3.555& 1.332 & 3.004 & 1.182& 4.820&2.166&2.901&0.953& 4.987 &2.374 &3.087&1.360&3.234&1.436&2.256&0.898\\
  \hline
  DTO&4.388& 1.958 & 4.839 &1.501&3.536&1.354&6.130&1.985&4.237&1.771&5.392&1.628&4.244&2.019&6.251&1.339\\
  \hline
  MA& 3.315&5.764&2.004&3.103&3.372&6.029&3.0886&4.685&2.562&2.356&2.476&4.831&2.480&3.243&2.985&4.708\\
  \hline
 \end{tabular}\\
    \begin{tabular}{|p{0.75cm}|p{3.71cm}|p{3.705cm}|p{3.705cm}|p{3.705cm}|}
      \hline
      Metric&No task&Visual task ($p<0.001$)***&Manual task ($p=0.02$)*& Visual-Manual task ($p=0.003$**)\\
     \end{tabular}\\
     \begin{tabular}{|p{0.75cm}|p{1.64cm}|p{1.63cm}|p{1.64cm}|p{1.64cm}|p{1.63cm}|p{1.64cm}|p{1.63cm}|p{1.64cm}|p{1.63cm}}
      \hline
      &No haptic&Haptic&No haptic& Haptic&No haptic& Haptic&No haptic& Haptic\\
     \end{tabular}\\
     \begin{tabular}{|p{.75cm}|p{0.6cm}|p{0.6cm}|p{0.6cm}|p{0.6cm}|p{0.6cm}|p{0.6cm}|p{0.6cm}|p{0.6cm}|p{0.6cm}|p{0.6cm}|p{0.6cm}|p{0.6cm}|p{0.6cm}|p{0.6cm}|p{0.6cm}|p{0.6cm}|}
      \hline 
      &yes&no&yes&no&yes&no&yes&no&yes&no&yes&no&yes&no&yes&no\\
      \hline
      Cols. &15&8&9&14&20&3&6&17&18&5&9&14&15&8&4&19\\
      \hline
    \end{tabular}\\

\end{table*}

\begin{table*}
 \centering
 \scriptsize
 \caption{
 Statistical analysis results from comparing the reaction time (RT), lateral root mean square error (RMSE), minimum distance to obstacle (DTO) and maximum acceleration (MA) depending on the activation of the system.}
 \label{table:nohaptic_vs_haptic}

 \begin{tabular}{|p{10.94cm}|}
 \hline
  \textbf{No haptic vs Haptic} t-test ($\alpha$ =0.05) \\
 \end{tabular}
 \begin{tabular}{|p{.75cm}|p{2.03cm}|p{2.34cm}|p{2.03cm}|p{2.04cm}|}
  \hline
  Metric&No Task &Visual task&Manual task & visual-manual task\\
 \end{tabular}
 \begin{tabular}{|p{.75cm}|p{0.8cm}|p{0.8cm}|p{0.8cm}|p{1.1cm}|p{0.8cm}|p{0.8cm}|p{0.8cm}|p{0.8cm}|}
  \hline
  &\emph{t(23)}&\emph{p}&\emph{t(23)}&\emph{p}&\emph{t(23)}&\emph{p}&\emph{t(23)}&\emph{p}\\
  \hline
  RT& -0.978 & 0.342 & 2.278 & \textbf{0.036}&0.981&0.340&-0.685&0.502\\
  \hline
  RMSE& 0.550 & 1.000 &3.739&\textbf{0.002}**&2.6&\textbf{0.019}*&2.942&\textbf{0.009}**\\
  \hline
  DTO& -0.418 & 0.681 & -5.552 & \textbf{$<$ .001}*** & 1.388 & 0.183 & 2.570 & \textbf{0.020}* \\
  \hline
  MA&0.005&0.998&0.3891&0.766&0.991&0.121&0.046&0.964\\
  \hline
 \end{tabular}
\end{table*}

\begin{table}[t]
 \centering
 \scriptsize
 \caption{ANOVA and Cochran's Q analysis results regarding reaction time (RT), lateral root mean square error (RMSE), minimum distance to collision object (DTO), maximum acceleration (MA), and the number of collisions (Cols).}
  \label{table:anova_cochran}
    \begin{tabular}{|p{4.3cm}|}
  \hline
  ANOVA ($\alpha$ =0.05)\\
 \end{tabular}\\
 \begin{tabular}{|p{1.3cm}|p{2.54cm}|}
  \hline
  Metric&Haptic \\
 \end{tabular}\\
 \begin{tabular}{|p{1.3cm}|p{0.8cm}|p{1.3cm}|}
  \hline
  &\emph{F(3)}&\emph{p}\\
  \hline
  RT&  2.163& 0.101 \\
  \hline
  RMSE& 1.963 & 0.128 \\
  \hline
  DTO& 2.380 & 0.077  \\
  \hline
  MA&0.068&0.977\\
  \hline
 \end{tabular}\\

 \begin{tabular}{|p{4.3cm}|}
  Cochran's Q test ($\alpha$ =0.05)\\
  \hline
 \end{tabular}\\
 \begin{tabular}{|p{1.3cm}|p{2.54cm}|}
  \hline
  Metric&Haptic  \\
 \end{tabular}\\
 \begin{tabular}{|p{1.3cm}|p{0.8cm}|p{1.3cm}|}
  \hline
  &\emph{F(23)}&\emph{p}\\
  \hline
  Cols.&3.857&0.277 \\
  \hline
 \end{tabular}\\
\end{table}


\subsection{Reaction time}
As depicted in Tables~\ref{table:mean_std} and~\ref{table:nohaptic_vs_haptic}, when comparing all the tasks with each other and the potential benefit of using the system, results from the paired sample student t-test showed that there was a statistical significant relationship between the reaction time to start the obstacle avoidance maneuver when participants were performing the visual task  (1.358 s without haptic guidance vs 1.086 s with haptic guidance).

The results from the ANOVA test (Table~\ref{table:anova_cochran}) showed that there was not a statistically significant relationship between the reaction time to avoid an obstacle when participants received guidance through the steering wheel and performing a \gls{NDRT}.

\subsection{Lateral root mean square error}
The analysis of the driving performance resulted in a statistically significant relationship between the lateral \gls{RMSE} and the use of the system. Tables~\ref{table:mean_std} and and~\ref{table:nohaptic_vs_haptic} show that there were differences for the visual task (4.820 m without haptic guidance vs 2.901 m with haptic guidance), manual task (4.987 m without haptic guidance vs 3.087 m with haptic guidance) and visual-manual task (3.234 m without haptic guidance vs 2.256 m with haptic guidance). 

The results from the ANOVA statistic test  showed that there were no statistically significant differences between using the system after a take over request and the performance of secondary tasks in general (Table~\ref{table:anova_cochran}).   

\subsection{Minimum distance to obstacle}
The results from comparing the minimum distance to the sudden obstacle on the road for each secondary task in scenarios without haptic guidance compared with haptic guidance, showed statistically significant differences when the participants were performing the visual task (3.536 m vs 6.130 m) and the visual-manual task (4.244 m vs 6.251 m). The effect of the use of the system on distance in the manual task was not statistically significant (see Tables~\ref{table:mean_std} and~\ref{table:nohaptic_vs_haptic}).

There were no statistically significant differences between the minimum distance to the obstacle on the road and the performance of \gls {NDRT} in general when the guiding system was active (Table~\ref{table:anova_cochran}).

\subsection{Maximum acceleration}

Results from the statistical tests to determine the effect of the system on the maximum acceleration after a TOR was triggered showed that there were no statistical significant differences related to the \gls{NDRT} performed. See Tables~\ref{table:mean_std}, ~\ref{table:nohaptic_vs_haptic} and \ref{table:anova_cochran}. 

\subsection{Obstacle collisions}

Statistical significant differences could be found  when aplying the McNemar's test in the visual task (20 collisions without haptic guidance compared with 6 collisions using the haptic guidance system), manual task (18 collision vs 9 collisions) and visual-manual task (15 collisions vs 4 collisions) (Table~\ref{table:mean_std}).

The Cochran Q test results obtained from comparing the number of collisions with the obstacle that suddenly appeared on the road, with the use of the guiding system showed that there were no statistical significant differences that related to the performance of secondary tasks (Table~\ref{table:anova_cochran}).  

.

\section{Summary of Findings and Discussion}
\label{sec:findings}

The data collected through the experiments performed in this study delivered interesting results regarding the effect of an assistance system to avoid an obstacle on the road after being the participants involved in \gls{NDRT}s, while the vehicle was driving in a level 3 automated mode and the driver was requested to take the control of the vehicle. Results showed that their reaction times to start the avoidance maneuver were slower after having  performed the visual task compared to the other \gls{NDRT}s. Apparently, the visual task required more attentional resources than the other tasks. This difference was however not statistically significant.

In most cases, the activation of the guiding system improved the participants' trajectories to avoid the obstacle. 
However, this was not always the case when no secondary task was performed. The performance of the driving task without \gls{NDRT} resulted in a faster reaction to the \gls{TOR}, and as a consequence a higher, safer distance to it. This was due to the fact that drivers started the avoidance maneuver faster and overrode the guidance system. However, the RMSE and MA values were still improved by the guidance system, even without being the driver involved in a secondary task.

In situations in which the drivers were engaged in the performance of secondary tasks, the time required for them to regain road situational awareness and obstacle avoidance readiness enabled the system to operate for a longer period of time without being overruled by the authority of the participants. As a result, the lateral errors during the avoidance maneuver were reduced. 

The activation of the haptic guidance system when \gls{NDRT}s were performed, also resulted in greater distances between the vehicle and the road obstacle, avoiding thus a possible collision. The same tendency could be seen in the  manual task, being however the increased values not statistically significant.

It is noteworthy to mention that we designed the scenario to emulate critical situations on purpose. Thus, we expected a high collision rate. However, in the scenarios in which the haptic guidance system was enabled, we observed a higher rate of participants that successfully avoided the obstacle. The haptic guidance system took the control of the vehicle faster than the driver, starting thus the avoidance maneuver sooner. 

These findings were in line with the findings reported in~\cite{9517293}, in which a higher rate of maneuvers that avoided the obstacle  occurred when the participants didn't have the control of the vehicle. 

Based on the analyses results, we reject the null hypotheses defined in section~\ref{sec:system_implementation_approach}, \textit{A}, \textit{B}, \textit{C} and \textit{E} and accept the alternative hypotheses \textit{H1} in those cases.
Maximum acceleration was not affected by the use of haptic guidance systems. Therefore, we accept the null hypothesis defined in section~\ref{sec:system_implementation_approach},  \textit{D}.

\section{Conclusion and Future Work}
\label{sec:conclusion}

In this work we studied the impact of haptic guidance systems on driver's ability to take back the control of a level 3 vehicle and avoid an obstacle on the road. To this end, we defined different scenarios in which a sudden event triggered a \gls{TOR}. We also exposed the participants to different secondary tasks to study the effect of visual and cognitive distraction on the driving performance. To this end, we defined as independent variables the haptic guidance system and the \gls{NDRT}.

Results showed that the use of haptic guidance systems positively affected driving performance after a take over request in a variety of situations that involved secondary tasks, being these systems therefore helpful to promote road safety. 

Future work will address different levels of haptic guidance that will be tailored to different scenarios. We will additionally explore realistic scenarios in field tests by relying on the use of a real vehicle.

\color{black}
\section*{ACKNOWLEDGMENT}
This work was partially supported by the Austrian Science Fund
(FWF), within the project ”Interaction of autonomous and
manually-controlled vehicles (IAMCV)”, number P 34485-N and the Austrian Ministry for Climate Action, Environment, Energy, Mobility, Innovation and Technology (BMK) Endowed Professorship for Sustainable Transport Logistics 4.0., IAV France S.A.S.U., IAV GmbH, Austrian Post AG and the UAS Technikum Wien.

\bibliographystyle{IEEEtran}
\bibliography{paper}
\end{document}